\documentclass{PoS}

\usepackage{subfig}
\usepackage{wrapfig}
\usepackage{slashed}

\title{Searches for Double Parton Scattering at the LHC}

\ShortTitle{Searches for Double Parton Scattering at the LHC}

\author{\speaker{Miroslav MYSKA}\\
        Institute of Physics, Academy of Sciences of the Czech Republic,\\
        Na Slovance 2, CZ - 18221, Prague, Czech Republic\\
        E-mail: \email{myskam@fzu.cz}}

\abstract{Searches for the kinematic selection criteria for
estimation of multiple parton interaction fraction in pp collisions
at $\sqrt{s}$ = 14 TeV are presented using the eikonalization of the
cross section implemented in the Herwig++ Monte Carlo event generator. $W^+$ boson
pair production is studied in the muon decay channel. Four types of
the main background processes are discussed and analyzed. These
include $W^+W^+jj$, $W^+Z$, $ZZ$, and $t\bar{t}$ productions of the
positively charged muon pairs. The double parton scattering
contribution to the requested di-muon final state is found to be
around 25 per cent with the production cross section of 0.94 fb.}

\FullConference{XXIst International Europhysics Conference on High Energy Physics\\
                 21-27 July 2011\\
                 Grenoble, Rh\^one-Alpes France}

\begin{document}

\section{Introduction}

The topic of the Multiple Parton Scattering (MPS) is discussed within the
hadron collider physics for many years. The main motivation for their
consideration is to explain the high activity in detectors in comparison
to the standard picture of Single Parton Scattering (SPS), especially for the low transverse momentum
multi-jet production. The parton model allows us to consider several independent soft or semi-hard parton-parton interactions
within the same hadron-hadron collision and so model the production of many
particles in the final state. All the three general-purpose
Monte Carlo event generators Herwig++ \cite{myska:Herwig}, Pythia8 \cite{myska:Pythia}, and Sherpa \cite{myska:Sherpa}, have implemented
their mechanisms in order to model MPS.

\section{$\mu^+\mu^+$ Production and SPS Background Estimation}

\begin{wrapfigure}{l}{4cm}
\begin{center}
\vspace{-1cm}
\includegraphics[width=4cm]{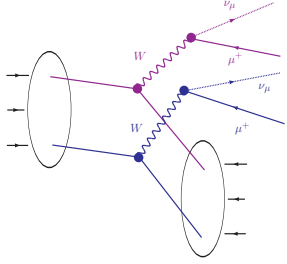}
\caption{Double parton scattering in proton-proton collision. A pair of $W^+$ bosons is produced via two independent parton annihilations.}
\label{myska:fig:WWdiagram}
\end{center}
\end{wrapfigure}

The vector boson pair production offers a very straightforward test
of the current stage of the MPS models. Despite the multi-jet
production, the double Drell-Yan process combines only limited
variations of quark-antiquark flavor pairs and can probe the
universality of the common normalization factor often called
effective cross section. The simultaneous production of two
positively charged $W$ bosons (see Fig.\ref{myska:fig:WWdiagram})
was chosen to be probed at the Monte Carlo level as the prediction
for the following measurement at the LHC experiments. The double
parton scattering signal was generated using the Herwig++ program
which is fully capable to generate this type of proton-proton events
within its underlying event model based on the eikonalization of the
cross section. In order to satisfy the inclusivity of this DPS
process in the number of parton sub-interactions in one event,
Herwig++ generates several $QCD2\rightarrow2$ sub-processes in
addition to two Drell-Yan annihilations according to the pre-sampled
distribution of the Poissonian shape \cite{myska:Herwig-eikonal}.

\begin{figure}[h!]
\begin{tabular}{ccc}
 \begin{minipage}{.3\hsize}
\includegraphics[width=5cm]{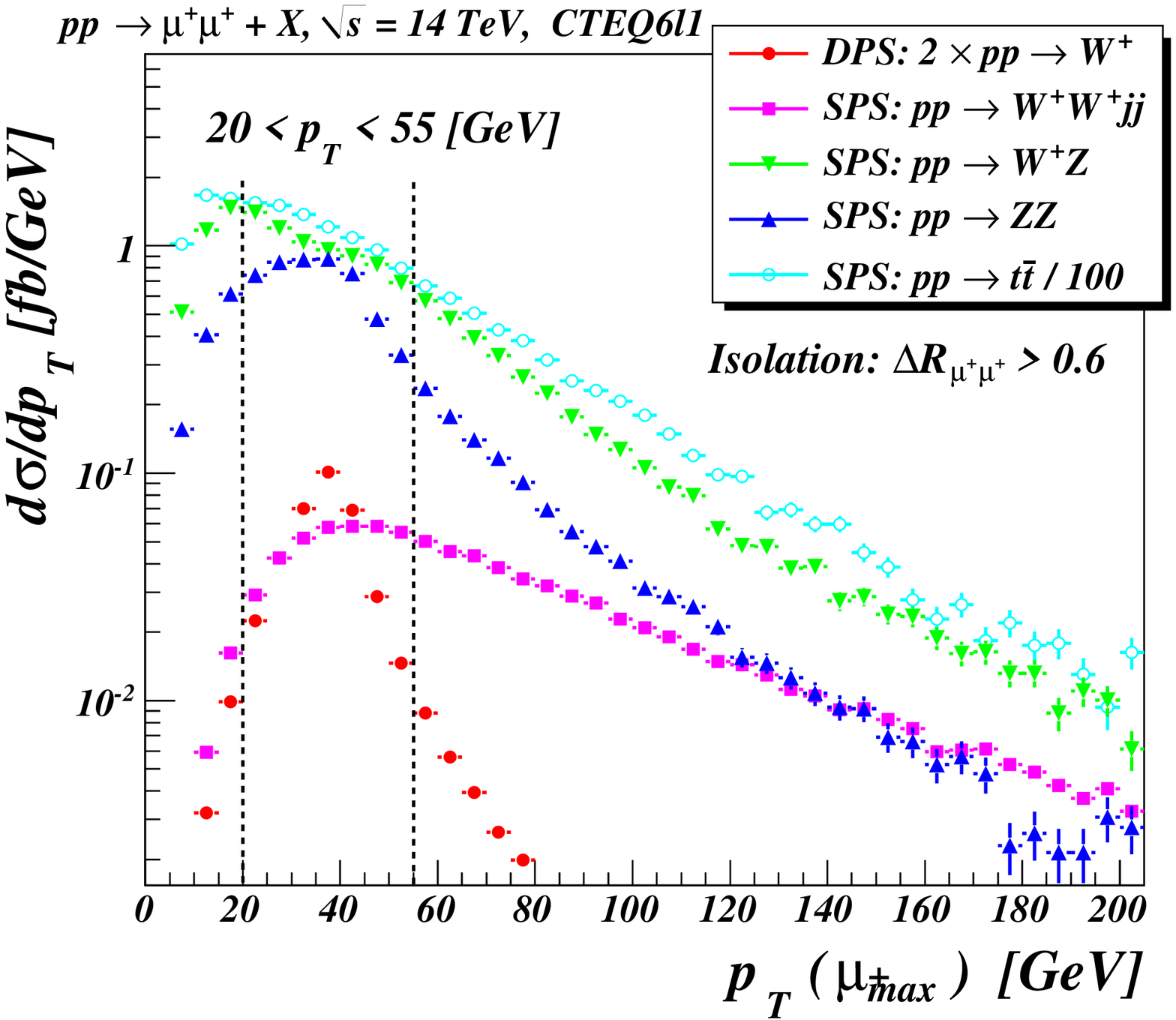}
\centering
{\small(a)}
 \end{minipage}
&
 \begin{minipage}{.3\hsize}
\includegraphics[width=5cm]{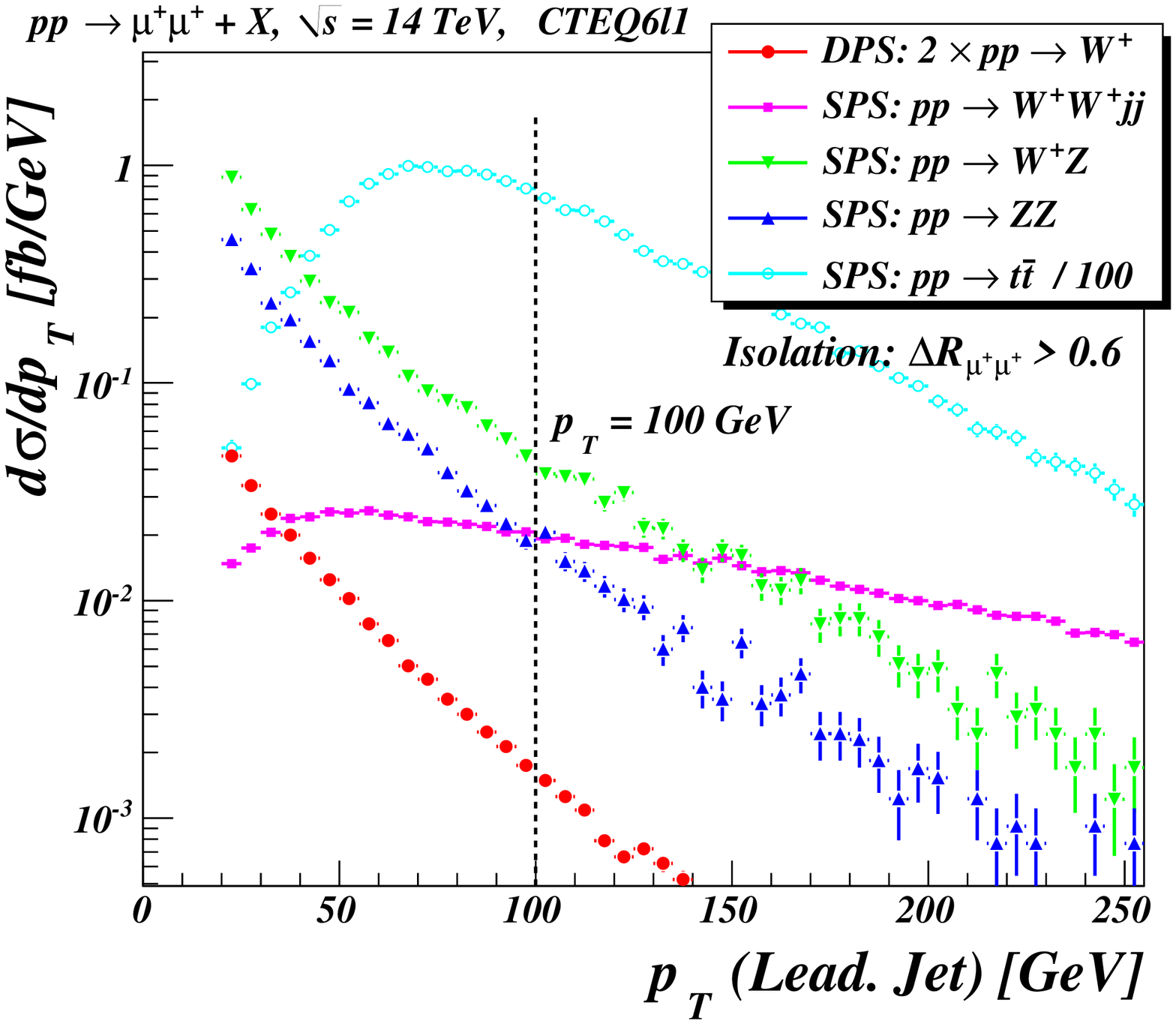}
\centering
{\small(b)}
 \end{minipage}
&
 \begin{minipage}{.3\hsize}
\includegraphics[width=5cm]{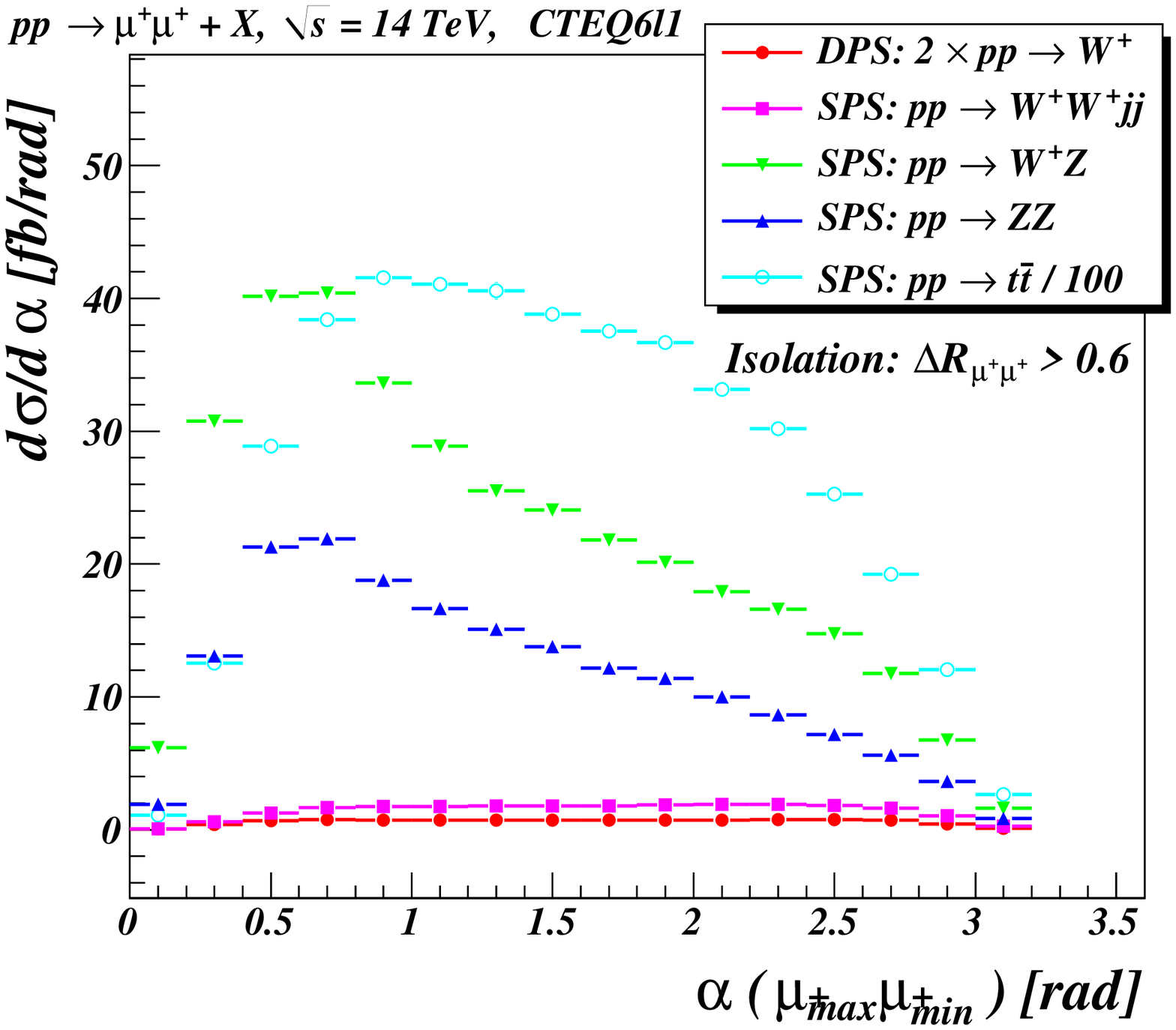}
\centering
{\small(c)}
\end{minipage}
\end{tabular}
\caption{Differential cross section as a function of (a) the transverse momentum of the hardest positively charged muon,
         (b) of the transverse momentum of the leading jet, (c) of the relative angle between the measured muons.}
\label{myska:fig:kinematics}
\end{figure}

Four types of physics background is studied in order to establish
the kinematical region, where the double parton scattering
production mechanism has a maximal significance against other
processes producing two same-sign muons in the final state. The
vector boson pair productions for three charge
combinations: $W^+W^+$, $W^+Z$, and $ZZ$, are prepared using the
MadGraph/MadEvent generator \cite{myska:Madgpraph}, where $Z$ stands
for both $\gamma$ and $Z$. Heavy flavor quark pair production is
generally the dominant source of the searched di-muons because of
the huge cross section, that leaves lot of space for the muon
radiation from short-living hadrons together with the heavy quark
decay to the hard muon. The $t\bar{t}$ production is studied using
the Herwig++ generator.

An example of the kinematical comparison among the individual
contributions to the di-muon final state is shown in
Fig.\ref{myska:fig:kinematics}. The $p_T$ spectrum of the two
hardest muons ($\mu^+_{max}$ and $\mu^+_{min}$) in the event is the
strongest selection tool together with the jet analysis. The final
result is also significantly dependent on the minimal $p_T$ of the
negatively charged muon ($\mu^-_{max}$) that can be measured within
the given detector acceptance. The final selection criteria are
summarized in Table \ref{myska:tab:final_selection}. Jets are
clustered using the $anti-k_t$ algorithm \cite{myska:antikt}
implemented in the FastJet package \cite{myska:Fastjet} with
$R=0.4$.

\begin{table}[h!]
\begin{center}
\begin{tabular}{|c c c| c c c| c c c|}
$|\eta(\mu)|$ & $<$ & $2.5$ & $|\eta(jet)|$ & $<$ & $4.5$ & $p_T(\mu^-_{max})$ & $<$ & $5~GeV$ \\
$p_T(\mu^+_{max})$ & $>$ & $20~GeV$ & $p_T(\mu^+_{min})$ & $>$ & $15~GeV$ & $\slashed{E}_T$ & $>$ &
$20~GeV$ \\
$p_T(\mu^+_{max})$ & $<$ & $50~GeV$ & $p_T(\mu^+_{min})$ & $<$ & $40~GeV$ & $\Delta R_{\mu^+ jet}$ & $>$ & $0.4$ \\
$M (\mu^+_{max}\mu^+_{min})$ & $>$ & $20~GeV$ & $M_T(\mu^+_{max}\mu^+_{min})$ & $>$ & $50~GeV$ & $\Delta R_{\mu^+_{max}\mu^+_{min}}$ & $>$ & $0.6$\\
$p_T(jet_{lead})$ & $<$ & $40~GeV$ & $p_T(jet_{sub-lead})$ & $<$ & $30~GeV$ & & &
\end{tabular}
\caption{Summary of the final selection criteria applied on Monte Carlo data.} \label{myska:tab:final_selection}
\end{center}
\end{table}

\vspace{-1cm}

\section{Summary and Conclusion}

Several mechanisms were studied in order to find the
kinematical region for the statistically reasonable measurement of
the double parton scattering. The final selection described in Table \ref{myska:tab:final_selection}
is considered as the main set of the selection criteria, where
signal-background ratio reaches 0.35. The result will also strongly depend on
the detector performance and on the trigger efficiencies.
Considering that the signal cross section in LO is 0.94 fb, the
integrated luminosity is required to be very large, of
$\mathcal{O}(100fb^{-1})$. However, the LHC could provide enough of
the necessary statistics within a few years of full energy
operation. Measurement of this process may bring a significant
insight into the hadron structure, especially for the transverse
parton distributions, and very valuable feedback for Monte Carlo
generators.

\end{document}